\documentclass[aps,pre,twocolumn,nofootinbib,floatfix]{revtex4}
\usepackage{amsmath}
\usepackage{epsfig}

\input{tcilatex}

\begin{document}

\title{Current noise of a quantum dot \textit{p-i-n} junction in a photonic
crystal}
\author{Y. N. Chen$^{1}$, D. S. Chuu$^{1}$, and T. Brandes$^{2}$}
\affiliation{$^{1}$Department of Electrophysics, National Chiao-Tung University, Hsinchu
300, Taiwan\\
$^{2}$School of Physics and Astronomy, The University of Manchester P.O. Box
88, Manchester, M60 1QD, U.K.}

\begin{abstract}
The shot-noise spectrum of a quantum dot \textit{p-i-n} junction embedded
inside a three-dimensional photonic crystal is investigated. Radiative decay
properties of quantum dot excitons can be obtained from the observation of
the current noise. The characteristic of the photonic band gap is revealed
in the current noise with discontinuous behavior. Applications of such a
device in entanglement generation and emission of single photons are pointed
out, and may be achieved with current technologies.

\end{abstract}

\maketitle

\address{$^{1}$Department of Electrophysics, National Chiao Tung University,
Hsinchu 30050, Taiwan\\
$^{2}$Department of Physics, UMIST, P.O. Box 88, Manchester, M60 1QD, U.K.}

\address{$^{1}$Department of Electrophysics, National Chiao Tung University,
Hsinchu 30050, Taiwan} 
\address{$^{2}$Department of Physics, UMIST, P.O. Box 88, Manchester, M60
1QD, U.K.}

\address{Department of Electrophysics, National Chiao Tung University,
Hsinchu 300, Taiwan}





Since Yablonovitch proposed the idea of photonic crystals (PCs) \cite{1},
optical properties in periodic dielectric structures have been investigated
intensively \cite{2}. Great attention has been focused on these materials
not only because of their potential applications in optical devices, but
also because of their ability to drastically alter the nature of the
propagation of light from a fundamental perspective \cite{3}. Among these,
modification of spontaneous emission is of particular interest.
Historically, the idea of controlling the spontaneous emission rate was
proposed by Purcell \cite{4}, and enhanced and inhibited spontaneous
emission rates for atomic systems were intensively investigated in the 1980s %
\cite{5} by using atoms passed through a cavity. In semiconductor systems,\
the electron-hole pair is naturally a candidate to examine spontaneous
emission, where modifications of the spontaneous emission rates of quantum
dot (QD) \cite{6} or quantum wire (QW) \cite{7} excitons inside the
microcavities have been observed experimentally.

Recently, the interest in measurements of shot noise in quantum transport
has risen owing to the possibility of extracting valuable information not
available in conventional dc transport experiments \cite{8}. With the
advances of fabrication technologies, it is now possible to embed QDs inside
a \textit{p-i-n} structure \cite{9}, such that the electron and hole can be
injected separately from opposite sides. This allows one to examine the
exciton dynamics in a QD via electrical currents \cite{10}. On the other
hand, it is also possible to embed semiconductor QDs in PCs \cite{11}, where
modified spontaneous emission of QD excitons is observed over large
frequency bandwidths.

In this work, we present non-equilibrium calculations for the quantum noise
properties of quantum dot excitons inside photonic crystals. We obtain the
current noise of QD excitons via the MacDonald formula \cite{12}, and find
that it reveals many of the characteristics of the photonic band gap (PBG).
Possible applications of such a device to the generation of entangled states
and the emission of single photons are also pointed out. 
\begin{figure}[th]
\includegraphics[width=7cm]{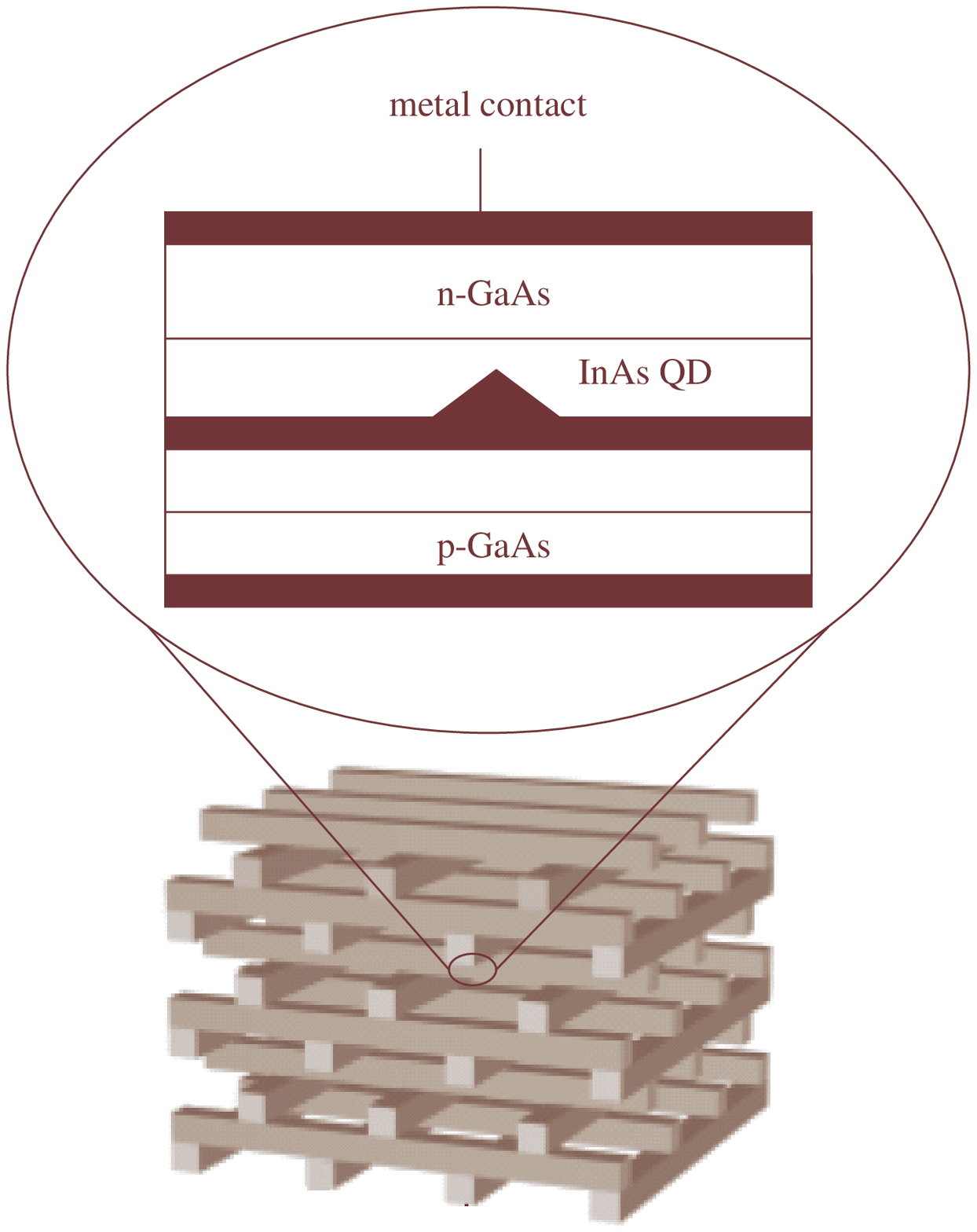}
\caption{{}Illustration of a QD inside a \textit{p-i-n} junction surrounded
by a three-dimensional PC.}
\end{figure}

\emph{The model.} --- We assume that a QD \textit{p-i-n} junction is
embedded in a three-dimensional PC. A possible structure is shown in Fig. 1.
Both the hole and electron reservoirs are assumed to be in thermal
equilibrium. For the physical phenomena we are interested in, the Fermi
level of the \textit{p(n)}-side hole (electron) is slightly lower (higher)
than the hole (electron) subband in the dot. After a hole is injected into
the hole subband in the QD, the \textit{n}-side electron can tunnel into the
exciton level because of the Coulomb interaction between the electron and
hole. Thus, we may introduce the three dot states: $\left| 0\right\rangle
=\left| 0,h\right\rangle $, $\left| \uparrow \right\rangle =\left|
e,h\right\rangle $, and $\left| \downarrow \right\rangle =\left|
0,0\right\rangle $, where $\left| 0,h\right\rangle $ means there is one hole
in the QD,\ $\left| e,h\right\rangle $ is the exciton state, and $\left|
0,0\right\rangle $ represents the ground state with no hole and electron in
the QD. One might argue that one can not neglect the state $\left|
e,0\right\rangle $ for real devices since the tunable variable is the
applied voltage. This can be resolved by fabricating a thicker barrier on
the electron side so that there is little chance for an electron to tunnel
in advance \cite{13}. Moreover, the charged exciton and biexcitons states
are also neglected in our calculations, which means a low injection limit is
required \cite{14}.

\emph{Derivation of Master equation.} ---We define the dot-operators $%
\overset{\wedge }{n_{\uparrow }}\equiv \left| \uparrow \right\rangle
\left\langle \uparrow \right| ,$ $\overset{\wedge }{n_{\downarrow }}\equiv
\left| \downarrow \right\rangle \left\langle \downarrow \right| ,$ $\overset{%
\wedge }{p}\equiv \left| \uparrow \right\rangle \left\langle \downarrow
\right| ,$ $\overset{\wedge }{s_{\uparrow }}\equiv \left| 0\right\rangle
\left\langle \uparrow \right| ,$ $\overset{\wedge }{s_{\downarrow }}\equiv
\left| 0\right\rangle \left\langle \downarrow \right| $. The total
Hamiltonian $H$ of the system consists of three parts: $H_{0}$ [dot, photon
bath $H_{p}$, and the electron (hole) reservoirs $H_{res}$], $H_{T}$
(dot-photon coupling), and the dot-reservoir coupling $H_{V}$:

\begin{eqnarray}
H &=&H_{0}+H_{T}+H_{V}  \notag \\
H_{0} &=&\varepsilon _{\uparrow }\overset{\wedge }{n_{\uparrow }}%
+\varepsilon _{\downarrow }\overset{\wedge }{n_{\downarrow }}+H_{p}+H_{res} 
\notag \\
H_{T} &=&\sum_{k}D_{k}b_{k}^{\dagger }\overset{\wedge }{p}+D_{k}^{\ast }b_{k}%
\overset{\wedge }{p}^{\dagger }=\overset{\wedge }{p}X+\overset{\wedge }{p}%
^{\dagger }X^{\dagger }  \notag \\
H_{p} &=&\sum_{k}\omega _{k}b_{k}^{\dagger }b_{k}  \notag \\
H_{V} &=&\sum_{\mathbf{q}}(V_{\mathbf{q}}c_{\mathbf{q}}^{\dagger }\overset{%
\wedge }{s_{\uparrow }}+W_{\mathbf{q}}d_{\mathbf{q}}^{\dagger }\overset{%
\wedge }{s_{\downarrow }}+c.c.)  \notag \\
H_{res} &=&\sum_{\mathbf{q}}\varepsilon _{\mathbf{q}}^{\uparrow }c_{\mathbf{q%
}}^{\dagger }c_{\mathbf{q}}+\sum_{\mathbf{q}}\varepsilon _{\mathbf{q}%
}^{\downarrow }d_{\mathbf{q}}^{\dagger }d_{\mathbf{q}}.
\end{eqnarray}%
In the above equation, $D_{k}=i\hbar \mathbf{\epsilon \cdot \mu }\sqrt{%
\omega _{\mathbf{k}}/(2\epsilon _{0}\hbar V)}$ is the dipole coupling
strength with $\mathbf{\epsilon }$ and $\mathbf{\mu }$ being the
polarization vector of the photon and the dipole moment of the exciton,
respectively. $b_{k}$ is the photon operator, $X=\sum_{k}D_{k}b_{k}^{\dagger
}$ ,$\ $and $c_{\mathbf{q}}$ and $d_{\mathbf{q}}$ denote the electron
operators in the left and right reservoirs, respectively.

The couplings to the electron and hole reservoirs are given by the standard
tunnel Hamiltonian $H_{V},$ where $V_{\mathbf{q}}$ and $W_{\mathbf{q}}$
couple the channels $\mathbf{q}$ of the electron and the hole reservoirs. If
the couplings to the electron and the hole reservoirs are weak, it is
reasonable to assume that the standard Born-Markov approximation with
respect to these couplings is valid. In this case, one can derive a master
equation from the exact time-evolution of the system. The equations of
motion can be expressed as (cf. \cite{15})

\begin{eqnarray}
\frac{\partial }{\partial t}\overset{\wedge }{\left\langle n_{\uparrow
}\right\rangle }_{t} &=&-\int dt^{\prime }[C(t-t^{\prime })+C^{\ast
}(t-t^{\prime })]\overset{\wedge }{\left\langle n_{\uparrow }\right\rangle }%
_{t^{\prime }}  \notag \\
&&+\Gamma _{L}[1-\overset{\wedge }{\left\langle n_{_{\uparrow
}}\right\rangle }_{t}-\overset{\wedge }{\left\langle n_{\downarrow
}\right\rangle }_{t}]
\end{eqnarray}

\begin{equation*}
\frac{\partial }{\partial t}\overset{\wedge }{\left\langle n_{\downarrow
}\right\rangle }_{t}=\int dt^{\prime }[C(t-t^{\prime })+C^{\ast
}(t-t^{\prime })]\overset{\wedge }{\left\langle n_{\uparrow }\right\rangle }%
_{t^{\prime }}-\Gamma _{R}\overset{\wedge }{\left\langle n_{\downarrow
}\right\rangle }_{t}]
\end{equation*}

\begin{equation*}
\frac{\partial }{\partial t}\overset{\wedge }{\left\langle p\right\rangle }%
_{t}=-\frac{1}{2}\int dt^{\prime }[C(t-t^{\prime })+C^{\ast }(t-t^{\prime })]%
\overset{\wedge }{\left\langle p\right\rangle }_{t^{\prime }}-\frac{\Gamma
_{R}}{2}\overset{\wedge }{\left\langle p\right\rangle }_{t},
\end{equation*}%
where $\Gamma _{L}$ $=2\pi \sum_{\mathbf{q}}V_{\mathbf{q}}^{2}\delta
(\varepsilon _{\uparrow }-\varepsilon _{\mathbf{q}}^{\uparrow })$ , $\Gamma
_{R}=2\pi \sum_{\mathbf{q}}W_{\mathbf{q}}^{2}\delta (\varepsilon
_{\downarrow }-\varepsilon _{\mathbf{q}}^{\downarrow })$, and $\varepsilon
=\hbar \omega _{0}=\varepsilon _{\uparrow }-\varepsilon _{\downarrow }$ is
the energy gap of the QD exciton. Here, $C(t-t^{\prime })$ $\equiv
\left\langle X_{t}X_{t^{\prime }}^{\dagger }\right\rangle _{0}$ is the
photon correlation function, and depends on the time interval only. We can
now define the Laplace transformation for real $z,$

\begin{eqnarray}
C_{\varepsilon }(z) &\equiv &\int_{0}^{\infty }dte^{-zt}e^{i\varepsilon
t}C(t)  \notag \\
n_{\uparrow }(z) &\equiv &\int_{0}^{\infty }dte^{-zt}\overset{\wedge }{%
\left\langle n_{\uparrow }\right\rangle }_{t}\text{ \ }etc.,\text{ }z>0
\end{eqnarray}%
and transform the whole equations of motion into $z$-space,

\begin{eqnarray}
n_{\uparrow }(z) &=&-(C_{\varepsilon }(z)+C_{\varepsilon }^{\ast
}(z))n_{\uparrow }(z)/z  \notag \\
&&+\frac{\Gamma _{L}}{z}(1/z-n_{\uparrow }(z)-n_{\downarrow }(z))  \notag \\
n_{\downarrow }(z) &=&(C_{\varepsilon }(z)+C_{\varepsilon }^{\ast
}(z))n_{\downarrow }(z)/z-\frac{\Gamma _{R}}{z}n_{\downarrow }(z)  \notag \\
p(z) &=&-\frac{1}{2}(C_{\varepsilon }(z)+C_{\varepsilon }^{\ast }(z))p(z)/z-%
\frac{\Gamma _{R}}{2z}p(z).
\end{eqnarray}%
These equations can then be solved algebraically, and the tunnel current
from the hole- or electron-side barrier

\begin{equation}
\overset{\wedge }{I}_{R}=-e\Gamma _{R}\overset{\wedge }{\left\langle
n_{\downarrow }\right\rangle }_{t},\text{ \ }\overset{\wedge }{I}%
_{L}=-e\Gamma _{L}[1-\overset{\wedge }{\left\langle n_{\uparrow
}\right\rangle }_{t}-\overset{\wedge }{\left\langle n_{\downarrow
}\right\rangle }_{t}]
\end{equation}%
can in principle be obtained by performing the inverse Laplace
transformation on Eqs. (4). Depending on the complexity of the correlation
function $C(t-t^{\prime })$ in the time domain, this can be a formidable
task which can however be avoided if one directly seeks the quantum noise:

\emph{Shot noise spectrum.} --- In a quantum conductor in nonequilibrium,
electronic current noise originates from the dynamical fluctuations of the
current around its average. To study correlations between carriers, we
relate the exciton dynamics with the hole reservoir operators by introducing
the degree of freedom $n$ as the number of holes that have tunneled through
the hole-side barrier \cite{16} and write 
\begin{gather}
\overset{\cdot }{n}_{0}^{(n)}(t)=-\Gamma _{L}n_{0}^{(n)}(t)+\Gamma
_{R}n_{\downarrow }^{(n-1)}(t),  \notag \\
\overset{\cdot }{n}_{\uparrow }^{(n)}(t)+\overset{\cdot }{n}_{\downarrow
}^{(n)}(t)=(\Gamma _{L}-\Gamma _{R})n_{0}^{(n)}(t).
\end{gather}%
Eqs. (6) allow us to calculate the particle current and the noise spectrum
from $P_{n}(t)=n_{0}^{(n)}(t)+n_{\uparrow }^{(n)}(t)+n_{\downarrow
}^{(n)}(t) $ which gives the total probability of finding $n$ electrons in
the collector by time $t$. In particular, the noise spectrum $S_{I_{R}}$ can
be calculated via the MacDonald formula \cite{12,17}, 
\begin{equation}
S_{I_{R}}(\omega )=2\omega e^{2}\int_{0}^{\infty }dt\sin (\omega t)\frac{d}{%
dt}[\left\langle n^{2}(t)\right\rangle -(t\left\langle I\right\rangle )^{2}],
\end{equation}%
where $\frac{d}{dt}\left\langle n^{2}(t)\right\rangle =\sum_{n}n^{2}\overset{%
\cdot }{P_{n}}(t)$. Solving Eqs. (6) and (4), we obtain 
\begin{equation}
S_{I_{R}}(\omega )=2eI\{1+\Gamma _{R}[\hat{n}_{\downarrow }(z=-i\omega )+%
\hat{n}_{\downarrow }(z=i\omega )]\}.
\end{equation}%
In the zero-frequency limit, Eq. (6) reduces to 
\begin{equation}
S_{I_{R}}(\omega =0)=2eI\{1+2\Gamma _{R}\frac{d}{dz}[z\hat{n}_{\downarrow
}(z)]_{z=0}\}.
\end{equation}%
As can be seen, there is no need to evaluate the correlation function $%
C(t-t^{\prime })$ in the time domain such that all one has to do is to solve
Eq. (4) in $z$-space.

\emph{Results and discussions} --- The above derivation shows that the noise
spectrum of the QD excitons depends strongly on $C_{\varepsilon }(z)$. Let
us now turn our attention to the spontaneous emission of a QD exciton in a
three dimensional PC, where the vacuum dispersion relation is strongly
modified: an anisotropic band-gap structure is formed on the surface of the
first Brillouin zone in the reciprocal lattice space. In general, the band
edge is associated with a finite collection of symmetrically placed points $%
\mathbf{k}_{0}^{i}$ leading to a three-dimensional band structure \cite{3}.
In our study, the transition energy of the QD exciton is assumed to be near
the band edge $\omega _{c}$. The dispersion relation for those wave vectors $%
\mathbf{k}$ whose directions are near one of the $\mathbf{k}_{0}^{i}$ can be
expressed approximately by $\omega _{\mathbf{k}}=$\ $\omega _{c}+A\left| 
\mathbf{k-k}_{0}^{i}\right| ^{2}$, where $A$ is a model dependent constant. %
\cite{18} Thus, the correlation function $C_{\varepsilon }(z)=\sum_{\mathbf{k%
}}\left| gD_{k}\right| ^{2}/[z+i(\omega _{\mathbf{k}}-\omega _{0})]$ can be
calculated around the directions of each $\mathbf{k}_{0}^{i}$ separately,
and is given by 
\begin{equation}
C_{\varepsilon }(z)=\frac{-i\omega _{0}^{2}\beta ^{3/2}}{\sqrt{\omega _{c}}+%
\sqrt{-iz-(\omega _{0}-\omega _{c})}},
\end{equation}%
with $\beta ^{3/2}=d^{2}\sum_{i}\sin ^{2}\theta _{i}/8\pi \epsilon _{0}\hbar
A^{3/2}$ \cite{19}. Here, $\hbar \omega _{0}$ is the transition energy of
the QD exciton, $d$ is the magnitude of the dipole moment, and $\theta _{i}$
is the angle between the dipole vector of the exciton and the $i$th $\mathbf{%
k}_{0}^{i}$. 
\begin{figure}[th]
\includegraphics[width=7.5cm]{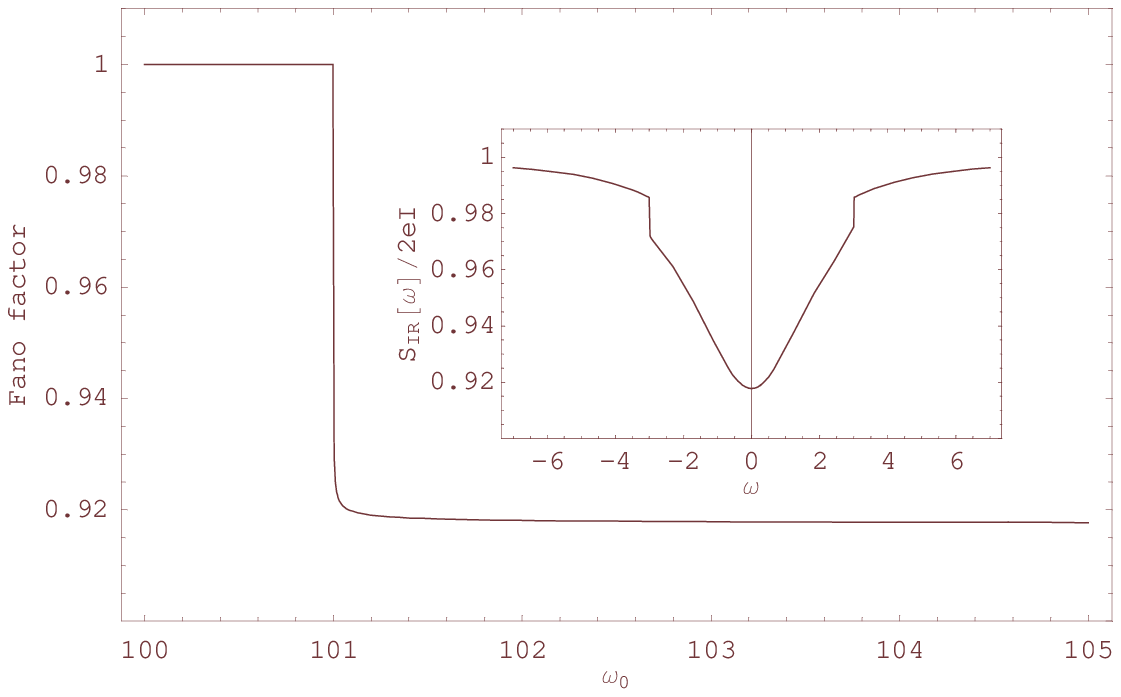}
\caption{{}Current noise (Fano factor) of QD excitons in a one-band PC as a
function of the exciton band gap $\protect\omega _{0}$. \ The PBG frequency $%
\protect\omega _{c}$ is set equal to $101\protect\beta $. The inset shows
frequency-dependent noise, in which $\protect\omega _{0}$ is fixed to $104%
\protect\beta $. }
\end{figure}

The shot-noise spectrum of QD excitons inside a PC is displayed in Fig. 2,
where the tunneling rates, $\Gamma _{L}$ and $\Gamma _{R}$, are assumed to
be equal to $0.1\beta $ and $\beta $, respectively. We see that the Fano
factor ($F\equiv S_{I_{R}}(\omega =0)/2e\langle I\rangle $) displays a
discontinuity as the exciton transition frequency is tuned across the PBG
frequency ($\omega _{c}=101\beta $). It also reflects the fact that below
the band edge frequency $\omega _{c}$, spontaneous emission of the QD
exciton is inhibited. To observe this experimentally, a DC electric field
(or magnetic field) could be applied in order to vary the band gap energy of
the QD exciton. Another way to examine the PBG frequency is to measure the
frequency-dependent noise as shown in the inset of Fig. 2, where the exciton
band gap is set equal to $104\beta $. As can be seen, discontinuities also
appear as $\omega $ is equal to the \textit{detuned} frequency between PBG
and QD exciton.

When the atomic resonant transition frequency is very close to the edge of
the band and the band gap is relatively large, the above one-band model is a
good approximation. If the band gap is narrow, one must consider both upper
and lower bands. For a three-dimensional anisotropic PC with point-group
symmetry, the dispersion relation near two band edges can be approximated as 
\begin{equation}
\omega _{k}=\QATOPD\{ . {\omega _{c_{1}}+C_{1}\left| \mathbf{k}-\mathbf{k}%
_{10}^{i}\right| \text{ \ }(\omega _{k}>\omega _{c_{1}}),}{\omega
_{c_{2}}-C_{2}\left| \mathbf{k}-\mathbf{k}_{20}^{j}\right| \text{ \ }(\omega
_{k}>\omega _{c_{1}}).}
\end{equation}%
Here, $\mathbf{k}_{10}^{i}$ and $\mathbf{k}_{20}^{j}$ are two finite
collections of symmetry related points, which are associated with the upper
and lower band edges \cite{20}, and $C_{1}$ and $C_{2}$ are model-dependent
constants. Following the derivation for the one-band PC, the correlation
function can now be written as 
\begin{equation}
C_{\varepsilon }(z)=\sum_{n=1}^{2}\frac{(-1)^{n}i\omega _{0}^{2}\beta
_{n}^{3/2}}{\sqrt{\omega _{c_{n}}}+\sqrt{(-1)^{n}\left[ iz+(\omega
_{0}-\omega _{c_{n}})\right] }},
\end{equation}%
where $\beta _{n}^{3/2}=d^{2}\sum_{i}\sin ^{2}\theta _{i}^{(n)}/8\pi
\epsilon _{0}\hbar C_{n}^{3/2}$ with the corresponding collections of angles 
$\theta _{i}^{(n)}$, $n=1,2$. 
\begin{figure}[th]
\includegraphics[width=7.5cm]{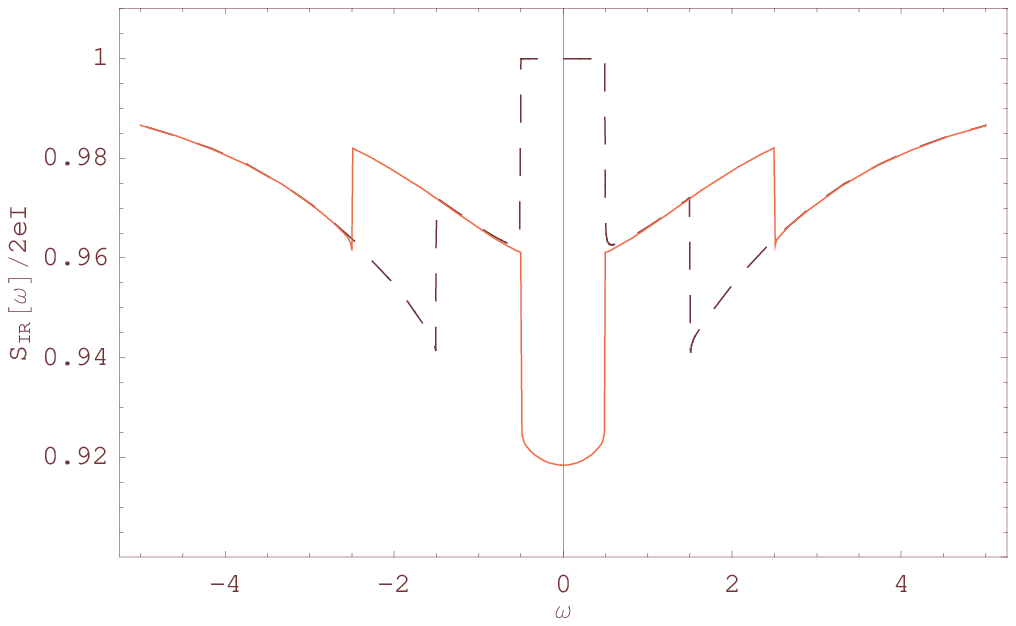}
\caption{{}Shot-noise spectrum of QD excitons in a two-band PC with $\protect%
\omega _{c_{1}}$ and $\protect\omega _{c_{2}}$ set equal to $101\protect%
\beta $ and $99\protect\beta $, respectively. To demonstrate the ability of
extracting information from the PC, the exciton band gap $\protect\omega %
_{0} $ in red and dashed curves is chosen as above $\protect\omega _{c_{2}}$
($\protect\omega _{0}=101.5\protect\beta $) and between the two band edge
frequencies ($\protect\omega _{0}=100.5\protect\beta $), respectively. }
\end{figure}

Fig. 3 illustrates the frequency-dependent noise of QD excitons embedded
inside a two-band PC. The two band edge frequencies $\omega _{c_{1}}$ and $%
\omega _{c_{2}}$ are set equal to $101\beta $ and $99\beta $, respectively.
There are three regimes for the choices of the exciton band gap: $\omega
_{0} $ $>\omega _{c_{1}}$, $\omega _{0}$ $<\omega _{c_{2}}$, and $\omega
_{c_{1}}>\omega _{0}>\omega _{c_{2}}$. When $\omega _{0}$ is tuned above the
upper band edge $\omega _{c_{1}}$ (or below the lower band edge $\omega
_{c_{2}}$), the QD exciton is allowed to decay, such that the shot noise
spectrum (red curve) is suppressed in the range of $\left| \omega \right|
<\left| \omega _{0}-\omega _{c_{1}}\right| $. On the other hand, however, if 
$\omega _{0}$ is between the two band edges, spontaneous emission is
inhibited. As shown by the dashed curve, the current noise in the central
region is increased with its value equal to unity. Similar to the one-band
PC, the curves of the shot noise spectrum reveal two discontinuities at $%
\left| \omega \right| =\left| \omega _{0}-\omega _{c_{1}}\right| $ or $%
\left| \omega _{0}-\omega _{c_{2}}\right| $, demonstrating the possibility
to extract information from a PC by the current noise.

A few remarks about the application of the QDs inside a PC should be
mentioned here. As is known, controlling the propagation of light
(waveguide) is one of the optoelectronic applications of PCs \cite{21}. By
controlling the exciton band gap $\omega _{0}$ across the PBG frequency with
appropriate tunneling rates of the electron and hole, one may achieve the
emission of a single photon at predetermined times and directions
(waveguides) \cite{22}, which are important for the field of quantum
information technology. Furthermore, it has been\ demonstrated recently that
a precise spatial and spectral overlap between a single self-assembled
quantum dot and a photonic crystal membrane nanocavity can be implemented by
a deterministic approach. \cite{23} One of the immediate applications is the
coupling of two QDs to a single common cavity mode. \cite{24} Therefore, if
two QD \textit{p-i-n} junctions can also be incorporated inside a PC (and on
the way of light propagation), the cavity-like effect may be used to create
an entangled state between two QD excitons with remote separation \cite{13}.
The obvious advantages then would be a suppression of decoherence of the
entangled state by the PBG, and the observation of the enhanced shot noise
could serve in order to identify the entangled state. \cite{10}

In summary, we have derived the non-equilibrium current noise of QD excitons
incorporated in a \textit{p-i-n} junction surrounded by a one-band or
two-band PC. We found that characteristic features of the PBG can be
obtained from the shot noise spectrum. Generalizations to other types of PCs
are expected to be relatively straightforward, which makes QD \textit{p-i-n}
junctions good detectors of quantum noise \cite{25}.

This work is supported partially by the National Science Council, Taiwan
under the grant numbers NSC 94-2112-M-009-019 and NSC 94-2120-M-009-002.


\begin{thebibliography}{99}
\bibitem{16} Actually, the total current noise should be expressed in terms
of the spectra of particle currents and the charge noise spectrum: $%
S_{I}(\omega )=aS_{I_{L}}(\omega )+bS_{I_{R}}(\omega )-ab\omega
^{2}S_{Q}(\omega )$, where $a$ and $b$ are capacitance coefficient ($a+b=1$)
of the junctions. Since we have assumed a highly asymmetric set-up ($a<<b$),
it is plausible to consider the hole-side spectra $S_{I_{R}}(\omega )$ only.
\bibitem{18} In fact, $A$ is dependent on the structures of the photonic
crystals. For a simple three-dimensional periodic dielectric, $A$ can be
approximated as $A\simeq \omega _{c}/(\mathbf{k}_{0}^{i})^{2}$, where $%
\mathbf{k}_{0}^{i}=\pi /L$ with $L$ being the lattice constant of the
photonic crystal. \cite{18} In order not to lose the generality, we let $A$
being a constant, such that $\beta $ becomes the unit for the numerical
calculations.
\end{thebibliography}
\end{document}